\documentclass[aps,prl,twocolumn,showpacs,superscriptaddress]{revtex4-1}

\usepackage[T1]{fontenc}
\usepackage[latin9]{inputenc}
\usepackage{color}
\usepackage{url}
\usepackage{textcomp}
\usepackage{amsmath}
\usepackage{amssymb}
\usepackage{graphicx}
\usepackage{natbib}
\usepackage[unicode=true,pdfusetitle,
 bookmarks=true,bookmarksnumbered=true,bookmarksopen=false,
 breaklinks=true,pdfborder={0 0 1},backref=false,colorlinks=true]
 {hyperref}
\hypersetup{
 pdfborderstyle=,linkcolor=red,citecolor=blue,urlcolor=blue}
 \usepackage[normalem]{ulem}

\makeatletter

\hypersetup{
 pdfborderstyle=,linkcolor=red,citecolor=blue,urlcolor=blue}



\begin{document}
\title{Signatures of a spin-$\frac{1}{2}$ cooperative paramagnet in the diluted triangular lattice of
Y$_{2}$CuTiO$_{6}$}
\author{S. Kundu}
\thanks{Equal contribution authors}
\affiliation{Department of Physics, Indian Institute of Technology Bombay, Powai,
Mumbai 400076, India}
\author{Akmal Hossain}
\thanks{Equal contribution authors}
\affiliation{Solid State and Structural Chemistry Unit, Indian Institute of Science,
Bengaluru 560012, India}
\author{Pranava Keerthi S}
\affiliation{Solid State and Structural Chemistry Unit, Indian Institute of Science,
Bengaluru 560012, India}
\author{Ranjan Das}
\affiliation{Solid State and Structural Chemistry Unit, Indian Institute of Science,
Bengaluru 560012, India}
\author{M. Baenitz}
\affiliation{Max Planck Institute for Chemical Physics of Solids, 01187 Dresden,
Germany}
\author{Peter J. Baker}
\affiliation{ISIS Pulsed Neutron and Muon Source, STFC Rutherford Appleton Laboratory,
Harwell Campus, Didcot, Oxfordshire OX110QX, UK}
\author{Jean-Christophe Orain}
\affiliation{Paul Scherrer Institute, Bulk MUSR group, LMU 5232 Villigen PSI, Switzerland}
\author{D. C. Joshi}
\affiliation{Department of Engineering Sciences, Uppsala University, Box 534, SE-751 21 Uppsala, Sweden}
\author{Roland Mathieu}
\affiliation{Department of Engineering Sciences, Uppsala University, Box 534, SE-751 21 Uppsala, Sweden}
\author{Priya Mahadevan}
\affiliation{S. N. Bose National Center for Basic Sciences, Block-JD, Salt lake, Kolkata-700106, India}
\author{Sumiran Pujari}
\affiliation{Department of Physics, Indian Institute of Technology Bombay, Powai,
Mumbai 400076, India}
\author{Subhro Bhattacharjee}
\affiliation{International Centre for Theoretical Sciences, Tata Institute of Fundamental
Research, Bengaluru 560089, India}
\author{A. V. Mahajan}
\email{mahajan@phy.iitb.ac.in}
\affiliation{Department of Physics, Indian Institute of Technology Bombay, Powai,
Mumbai 400076, India}
\author{D. D. Sarma}
\email{sarma@iisc.ac.in}
\affiliation{Solid State and Structural Chemistry Unit, Indian Institute of Science,
Bengaluru 560012, India}
\date{\today}
\begin{abstract}
We present a combination of thermodynamic and dynamic experimental signatures of a disorder driven dynamic cooperative paramagnet in a $50\%$ site diluted triangular lattice spin-$\frac{1}{2}$ system, Y$_{2}$CuTiO$_{6}$. Magnetic ordering and spin freezing are absent down to $50$ mK, far below the Curie Weiss scale of $\sim$$-134$ K. We observe scaling collapses of the magnetic field- and temperature-dependent magnetic heat capacity and magnetisation data, respectively, in conformity with expectations from the random singlet physics. Our experiments establish the suppression of any freezing scale, if at all present, by more than three orders of magnitude, opening a plethora of interesting possibilities such as {\it disorder-stabilized} long range quantum entangled ground states.

\end{abstract}
\maketitle


 Conventional wisdom suggests that structural disorder in magnetic insulators usually leads to random spin-spin exchanges, which, in turn, promotes spin freezing at low temperatures \cite{bhatt1982scaling,binder1986spin,nordblad2013competing}. There is, however, an interesting alternative where  quenched randomness may promote competing magnetic interactions and quantum fluctuations, thereby enhancing the possibility of realizing  quantum spin liquids (QSLs) \cite{balents2010spin}, as recently suggested for Ba$_{3}$CuSb$_{2}$O$_{9}$ \citep{Smerald2015} and Pr$_2$Zr$_2$O$_7$  \citep{Savary2017a}.

This raises several interesting and experimentally relevant questions -- {\it can structural disorder in magnetic insulators enhance quantum fluctuations and drive a magnetically ordered state (in  clean limit) to a quantum paramagnet}? What then is the nature of such a  paramagnet? Can such a state support non-trivial many-body entanglement and realise a disorder driven QSL?\cite{Kimchi2018a,furukawa2015quantum} This possibility of realizing QSLs \cite{potter2012quantum,voelker2001multiparticle,kim2006role} and associated novel superconductors \citep{baskaran2008resonating} arising from the interplay of disorder and interactions near the metal-insulator transition were explored theoretically in the context of doped semiconductors and boron doped diamond. On the experimental front, low temperature dynamic paramagnetism was observed in irradiation induced disordered organic magnet $\kappa$-(ET)$_2$Cu[N(CN)$_2$]Cl \cite{furukawa2015quantum}. However, the question of a disorder driven QSL is different from that of the effect of disorder in a clean QSL. In this latter case, the QSL is often stable to at least dilute local impurities albeit with interesting defect states \cite{willans2010disorder,willans2011site,kolezhuk2006theory,trousselet2011effects,petrova2014unpaired,sreejith2016vacancies}.

These issues are particularly pertinent for two dimensional spin-$\frac{1}{2}$ frustrated magnets where reduced dimensionality enhances quantum fluctuations and suppresses ordering tendencies. Thus, they serve as natural platforms to realize QSLs as in candidate materials-- Herbertsmithite  (ZnCu$_{3}$(OH)$_{6}$Cl$_{2}$) \citep{Mendels2007,Helton2007,Imai2008,Olariu2008,Vries2009,han2016correlated}  and $\mathrm{\kappa}$--(ET)$_{2}$Cu$_{2}$(CN)$_{3}$ \citep{Shimizu2003}. Notably, in Herbertsmithite, QSL is suggested to be stable to off-plane Cu$^{2+}$ magnetic disorder \citep{singh2010valence,Kimchi2018}.  Similarly, in three dimensional hyperkagome QSL candidate Na$_{4}$Ir$_{3}$O$_{8}$ \citep{Okamoto2007,Singh2013}, the coexistence of slow time-scales and signatures of quantum fluctuations have been observed \citep{Dally2014}.

\begin{figure}
\begin{centering}
\includegraphics[width=0.9\linewidth]{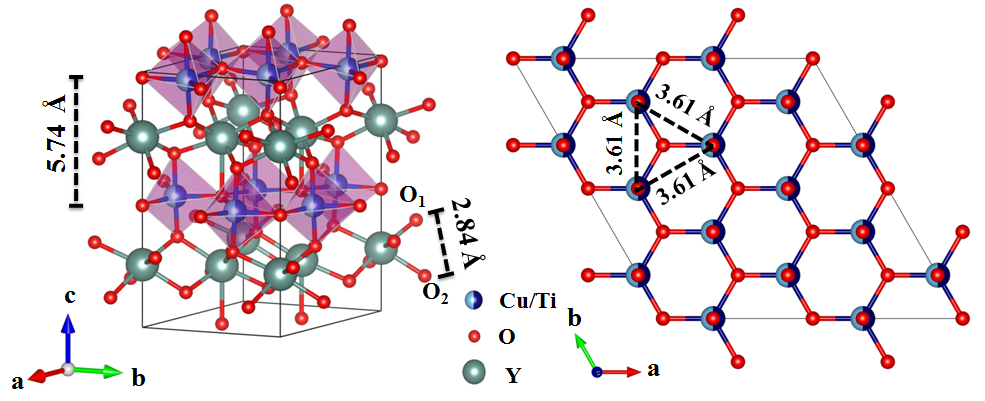}
\par\end{centering}
\centering{}
\caption{\label{fig:Double-perovskite-structure} Double perovskite structure
with one unit cell of Y$_{2}$CuTiO$_{6}$ and corner shared polyhedra
of ($\mathrm{Cu/Ti}$)O$_{5}$ connected via oxygen atoms in the $\mathit{ab}$-plane.
The $\mathrm{Cu^{2+}/Ti^{4+}}$ ions form edge-shared triangles.}
\end{figure}

In this paper, we report an experimental realization of a dynamic cooperative paramagnet in a spin-$\frac{1}{2}$ magnet on a site diluted triangular lattice $\mathrm{Y_{2}CuTiO_{6}}$ (YCTO) \citep{Choudhury2010,Floros2002}. This is established by the absence of any ordered or frozen magnetism down to the lowest accessible temperature (50 mK) despite substantial magnetic interactions, indicated by a large Curie-Weiss temperature ($\theta_{\mathrm{CW}} \sim$$-134$\,K) and disorder in the system. YCTO, therefore, is an example of a disordered triangular lattice magnet with 50:50 random mixture of spin-$\frac{1}{2}$ Cu$^{2+}$ atoms and non-magnetic Ti$^{4+}$ atoms, as shown in the left frame of Fig. \ref{fig:Double-perovskite-structure}. We observe a specific scaling behavior of thermodynamic quantities \citep{fisher1994random, bhatt1981scaling,bhatt1982scaling,Paalanen1988,dasgupta1980low} indicating a low temperature dynamic spin-$\frac{1}{2}$  paramagnet with possible formation of random singlets that survive down to the lowest accessible temperature \footnote{In the literature, there are at least two separate mentions of Random Singlet States in disordered quantum magnets. One is that of Ref. \citep{Kimchi2018a} with quenched disorder on which Ref. \citep{Kimchi2018} is based. For strong disorder, the picture is that of pinned short range valence bonds in the background, with topological spin-$\frac{1}{2}$ defects which experience power-law distributed interactions ($p(J)~\propto J^{-\gamma}$) as in Ref. \citep{Kimchi2018} leading to the low-energy phenomenology. The other scenario considered is the presence of strong exchange disorder in triangular lattices as in Ref. \citep{watanabe2014quantum,Shimokawa2015}}.

 In YCTO, each Cu/Ti site is surrounded by a triangular bipyramid of five oxygen atoms with three in the basal (a, b) plane and two along the c-axis. The resultant (Cu/Ti)O$_5$ units arranged in a triangular lattice in the (a,b) plane are well-separated by an intervening layer of non-magnetic Y$^{3+}$ ions along the c-axis with a large interlayer separation of $\sim$5.7 \AA.
 The exchange interaction strengths between nearest neighbor Cu atoms in the (a,b) plane ($J_{nn}$)
and the interlayer magnetic coupling ($J_c$) were calculated using density functional theory (see Supplemental Material (SM) \citep{SM-YCTO} for calculation details).
${J}_{nn}/k_B$ and $J_c/k_B$ were determined to be  $\sim -33.6$ K and $\sim -1.0$ K, respectively. These estimates lead to a calculated $\theta_{\mathrm{CW}}$ of $\sim -104$ K in reasonable agreement with the experimental value of $-134$ K. Larger than an order of magnitude anisotropy in magnetic interactions indicates
 two-dimensional triangular lattices of corner-shared (Cu/Ti)O$_5$ units, as shown in the right frame of Fig. \ref{fig:Double-perovskite-structure}, are coupled only weakly along the c-axis. Thus, we may think of the system approximately as
 one spin-$\frac{1}{2}$ at each Cu$^{2+}$ site sitting on an isotropic triangular lattice which is $50\%$ diluted with non-magnetic Ti$^{4+}$ ions. Experimentally no superlattice formation is observed in spite of the charge difference of the Cu$^{2+}$ and Ti$^{4+}$ ions  with the structural uniformity achieved only at a statistical level.
 We also note that superlattice formations on a triangular motif is geometrically frustrated at 50$\%$ dilution \footnote{The lowest electrostatic energy configuration space of the Ti$^{4+}$ and Cu$^{2+}$ ions is highly degenerate. This is easiest to see by mapping to frustrated classical Ising configurations on the triangular lattice. We also note that the system is not at the percolation threshold despite 50\% occupancy of the triangular lattice due to the presence of further neighbor magnetic exchange interactions, as discussed later in the text.}. Thus, YCTO, to a very good approximation, is a randomly 50$\%$ diluted spin-rotation invariant spin-$\frac{1}{2}$ magnet on a triangular lattice. Note that the random in-plane dilution and the spin-rotation symmetry are the two key differences between YCTO and the recently much investigated YbMgGaO$_{4}$ \citep{Kimchi2018a,li2015gapless}.

Polycrystalline samples of YCTO were synthesized by standard solid state reaction techniques. Details of the sample preparation and  measurements (magnetization, heat capacity, $^{89}$Y nuclear magnetic resonance (NMR), and muon spin relaxation ($\mu$SR)) are presented in Supplemental Material \citep{SM-YCTO}. From Rietveld refinements of x-ray diffraction data (see SM  \citep{SM-YCTO} for details), we confirmed that YCTO crystallizes in the non-centrosymmetric hexagonal structure  with the space group $\mathrm{\mathit{P}6_{3}\mathrm{\mathit{cm}}}$  \citep{Floros2002,Choudhury2010},
isostructural to  $\mathrm{LuMnO_{3}}$ \citep{Yakel1963}.

\begin{figure}
\centering{}
\includegraphics[scale=0.325]{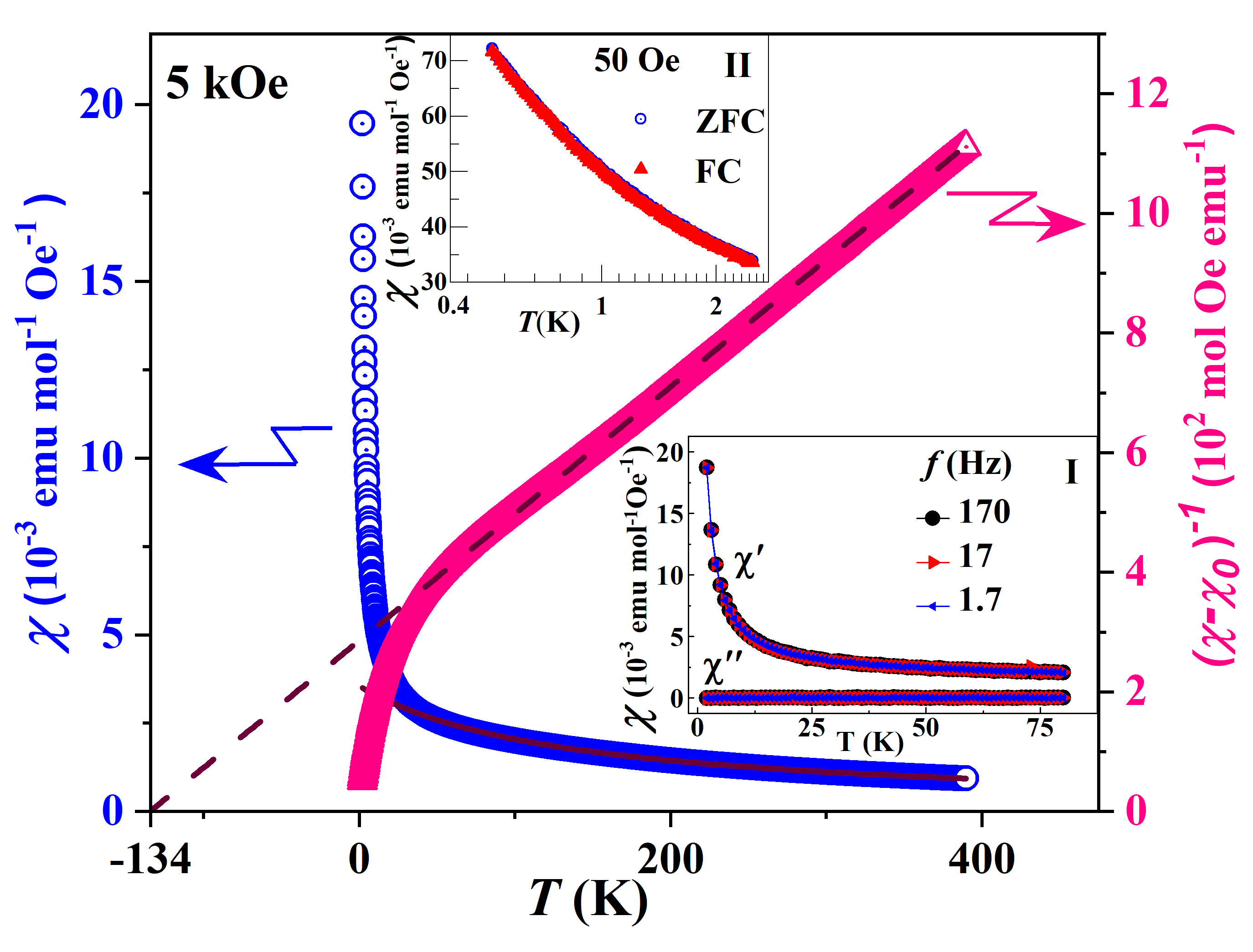}
\caption{\label{fig:susceptibility}The left y-axis shows $\mathit{\chi}(T)$ (open blue circles) of Y$_{2}$CuTiO$_{6}$ and the right $y$-axis shows the inverse susceptibility (open pink triangles) free from $\chi_{0}$. The Curie-Weiss fit is shown in the $T$- range 200-400 K with a solid line. The intercept on the $x$-axis gives $\theta_{CW}$ of about -134 K. The inset-I shows the ac susceptibility for different frequencies till 2 K and inset-II shows the absence of any bifurcation in the ZFC/FC data in 50 Oe down to 500 mK.}
\end{figure}

The dc susceptibility, $\mathit{\chi}$($\mathit{T}$) $=$ $M(T)/H$, as a function of $T$, is shown in Fig. \ref{fig:susceptibility}, where no signature of any magnetic ordering is seen down to 2 K. The divergence of $\chi$ at lowest temperatures is well-known to generally arise from the presence of a minor fraction ($\sim$2\% in the present case) of free spins in many such systems, as
detailed in SM \citep{SM-YCTO}.
A fitting of the high temperature (200-400 K) dc susceptibility to a Curie-Weiss form $\chi=\chi_{0}+C/(T-\theta_{\mathrm{CW}})$; where $\chi_{0}$, $C$ and $\theta_{\mathrm{CW}}$ are temperature independent Van Vleck paramagnetic and core diamagnetic susceptibilities, Curie constant, and the Curie-Weiss temperature, respectively, yields $\chi_{0}$ = 3.48$\times$10$^{-4}$ emu mol$^{-1}$ Oe$^{-1}$, $C$ = 0.47 emu K mol$^{-1}$ Oe$^{-1}$ and the $\theta_{\mathrm{CW}}$ = -134 K. The value of  $\theta_{\mathrm{CW}}$ is in agreement with earlier reports \citep{Choudhury2010,Singh2011} and the inferred moment of $\mu_{\mathrm{eff}}=1.94~\mu_{\mathrm{B}}$ is consistent with the expected value of $\sim$$1.9$ $\mu_{\mathrm{B}}$ for a spin $S = \frac{1}{2}$ Cu$^{2+}$ system with $\mathit{g}$ =2.2, as reported for many cuprates. Large and negative $\theta_{{\rm CW}}$, suggests substantial antiferromagnetic spin-spin interactions within each triangular layer.

The real and the imaginary parts of the ac susceptibility (see inset-I of Fig. \ref{fig:susceptibility}) did not show any indication of magnetic ordering. Further,  the lack of frequency dependence of the ac susceptibility over a rage of frequencies  indicates the absence of spin-freezing down to $2$ K. The lack of freezing despite extensive disorder  is re-emphasized by the absence of any irreversibility between the FC and ZFC magnetisation data at a low-field (50 Oe) (see inset-II of Fig. \ref{fig:susceptibility}), also ruling out any magnetic ordering down to 500 mK.  Thus, this system with substantial antiferromagnetic interactions on an essentially two-dimensional triangular lattice with  spin-$\frac{1}{2}$s and a relatively small spin-orbit coupling strength with no evidence of any magnetic ordering/freezing down to 500 mK offers a unique opportunity to probe a dynamic low temperature correlated paramagnetic phase that can possibly harbor intricate interplay of quantum fluctuations and disorder.

We extend the limit of the low temperature probe down to 50 mK with $\mu$SR experiments. Fig. \ref{fig:musr-lambda}a shows smooth exponential depolarisation of muon spins without any oscillations at all temperatures in absence of an external magnetic field. This shows that the fluctuation frequency of the local fields is much greater than the muSR frequency, ruling  out any static magnetic ordering down to 50 mK, though it cannot rule out a dynamic magnetic order \cite{PhysRevLett.96.127202}. We fit the muon spin polarization data to $e^{-\lambda t}$, as detailed in SM  \citep{SM-YCTO}, and obtain the muon spin relaxation rate,  $\lambda$, plotted as a function of the temperature in the inset of Fig. {\ref{fig:musr-lambda}a. We note that for a random internal static magnetic field with Gaussian distribution and zero mean, one expects a muon signal to decay as $e^{-at^2}$ with $a$ being a constant which is different from the $e^{-\lambda t}$ dependence seen in our experiments \cite{lacroix2011introduction}.

The initial increase in  $\lambda$  (inset of Fig.  \ref{fig:musr-lambda}a) on cooling indicates the expected slowing down of spin dynamics of Cu moments with lowering of  temperature down to about 2 K below which it essentially levels off at a value of $\lambda\approx 0.1~(\mu s)^{-1}$ down to $\sim 50$ mK. This indicates a dynamic low temperature state \citep{Li2016, Clark2013}.  If the observed muon depolarization arises from any static magnetic fields, $\lambda\approx0.1$ ($\mu$s)$^{-1}$ would suggest an estimate of the field ($\simeq$ $2\pi\lambda/\gamma_\mu$, with $\gamma_\mu$ being the muon gyromagnetic ratio) to be about 7.4 Oe. Then the muon spins can be decoupled from the static moments with the application of a longitudinal field about ten times this internal field \cite{lacroix2011introduction}. We have measured the muon polarization as a function of time at various temperatures and and applied longitudinal fields up to 2048 Oe-- about 270 times larger than 7.4 Oe, yet we did not see the total suppression of muon depolarization indicating existence of strong fluctuating local magnetic field in the system. A set of representative data obtained at 500 mK is shown in Fig.  {\ref{fig:musr-lambda}b. The corresponding decay constants, $\lambda$, are shown in the inset to Fig. \ref{fig:musr-lambda}b as a function of the applied field for different temperatures.
At all the measured temperatures $\lambda$ is nearly constant apart from a peak around 10 Oe which is presumably due to a quadrupolar level crossing resonance \citep{Kreitzman1986}  coming from the neighboring copper nuclei. This temperature independence suggests rapid spin fluctuations and the absence of any spin ordering or freezing down to at least 50 mK, despite sizable magnetic interactions.

\begin{figure}
\includegraphics[scale=0.10]{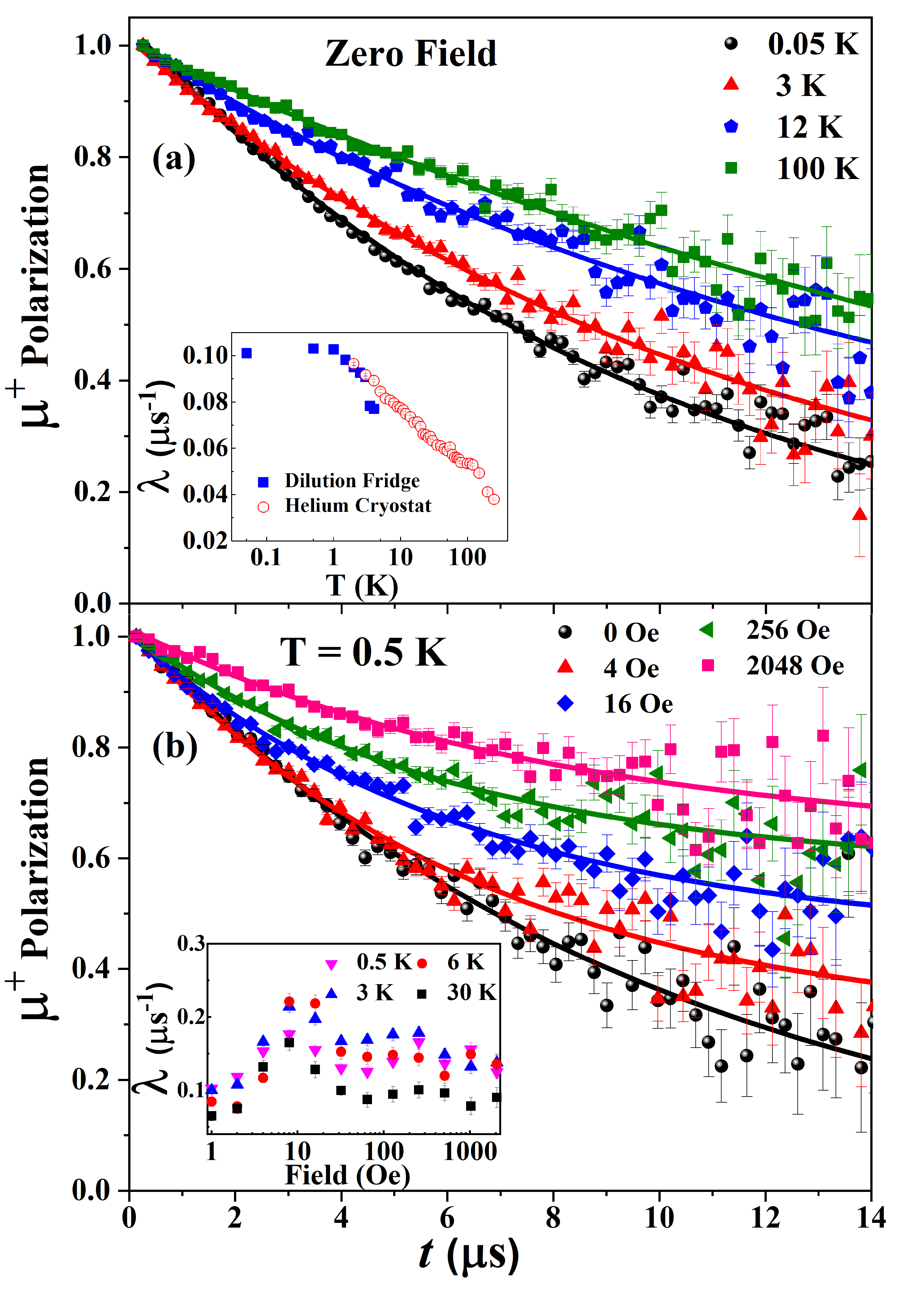}
\centering{}\caption{\label{fig:musr-lambda}Muon depolarization with time is shown (a) for various temperatures at zero fields. Solid lines are fits as described in the text. The inset shows the variation of the obtained muon relaxation rate with temperature, (b) for various longitudinal fields at 0.5 K and the corresponding inset shows the variation of the obtained muon relaxation rate with field.}
\end{figure}


This observation of the absence of spin-freezing or magnetic ordering down to $50$ mK then raises the interesting possibility of realizing disorder driven quantum paramagnetism in YCTO.
The lack of spin ordering/freezing till a very low temperature is also emphasized in the total heat capacity, $\mathit{\mathit{C}_{\mathrm{p}}\mathrm{(}T\mathrm{)}}$,
 (see SM \citep{SM-YCTO}) measured down to 350 mK for various magnetic fields.
The total contribution, $C_s$, of all spins to the specific heat is estimated by subtracting the lattice contribution, $C_{lat}$, from $C_p$. We find  $C_s/T \approx 0$ by 20 K (inset to Fig. \ref{fig:merged-heat-capacity}a). Spin entropy by integrating $C_s/T$  {\it vs.} $T$ up to 20 K is shown in Fig. \ref{fig:merged-heat-capacity}a for different applied magnetic fields. Clearly, this procedure accounts for only about 15\% of the total spin-entropy per mole of spin-$\frac{1}{2}$ Cu ions, evidencing a huge unquenched spin entropic content down to lowest temperatures (350 mK) complementing the muon relaxation data. We can further remove the contribution of free-spins from $C_s$ by subtracting the Schottky terms as detailed in SM \citep{SM-YCTO}, thereby defining $\mathit{C_{\mathrm{m}}}$. Inset to Fig. \ref{fig:merged-heat-capacity}b shows
the $T$-dependence of
$\mathit{C_{\mathrm{m}}}$ vs. $T$ at different fields on a log-log
scale for better clarity in the low temperature region. The data clearly
show that above $\sim$2 K, $\mathit{C_{m}}$ is independent of the applied magnetic
field.
Below 2\,K, $\mathit{C_{\mathrm{m}}}$ follows a
power law ($C_{\mathrm{m}}$ $\approx$ $\beta$$\mathit{T}$$^{1.4}$)
with temperature
indicating the presence of a large number of low
energy excitations. The regime of power-law behavior of $C_m$ shrinks with a decreasing magnetic field to a fraction of a Kelvin for a field of $1~{\rm T}$ suggesting that the low energy excitations directly couple to the magnetic field which is characteristic of random singlets with a distribution of bond energies \cite{Kimchi2018a,Kimchi2018}.

With the above results, we now turn to investigate the nature of the low temperature dynamic, correlated paramagnet.  The primary in-plane super-exchange between two nearest neighbour Cu$^{2+}$ atoms is mediated by the intermediate in-plane oxygen as is evident in  Fig. \ref{fig:Double-perovskite-structure}. Further neighbour exchanges either involve more intermediate O$^{2-}$ and in-plane Cu$^{2+}$/Ti$^{4+}$ or out-of-plane Y$^{3+}$-- hence expected to be suppressed similar to the inter-plane magnetic exchanges. This indicates that the magnetic physics of YCTO can be understood within a minimal model of diluted short-range antiferromagnet on a triangular lattice with a Hamiltonian : $H=\sum J_{\bf rr'} \eta_{\bf r}\eta_{\bf r'}{\bf S_r\cdot S_{r'}}$ where ${\bf S_r}$ are spin-$\frac{1}{2}$ operators on the triangular lattice sites, ${\bf r}$, and $\eta_{\bf r}=0(1)$ for a Ti$^{4+}$(Cu$^{2+}$)  site\cite{PhysRevB.82.134437,PhysRevB.74.144418}.  The distribution of Ti$^{4+}$ and Cu$^{2+}$ in 1:1 ratio is given by a probability distribution function, $\mathcal{P}[\{\eta_{\bf r}\}]$. $J_{\bf rr'}$ denotes short range antiferromagnetic interactions between the Cu$^{2+}$ ions and the randomness of the position of Cu$^{2+}$ lead to a distribution in $J_{\bf rr'}$ given by $\bar{\mathcal{P}}[\{J_{\bf rr'}\}]$ which is correlated with (but not necessarily same as) $\mathcal{P}[\{\eta_{\bf r}\}]$. The absence of signatures of formation of a super-lattice or any other structural anomalies in diffraction experiments suggests $\mathcal{P}[\{\eta_{\bf r}\}]$ to have weak correlations among different sites. Therefore, we expect that $\bar{\mathcal{P}}[\{ J_{\bf rr'}\}]$ has a relatively small width.

\begin{figure}

\includegraphics[width=0.9\linewidth]{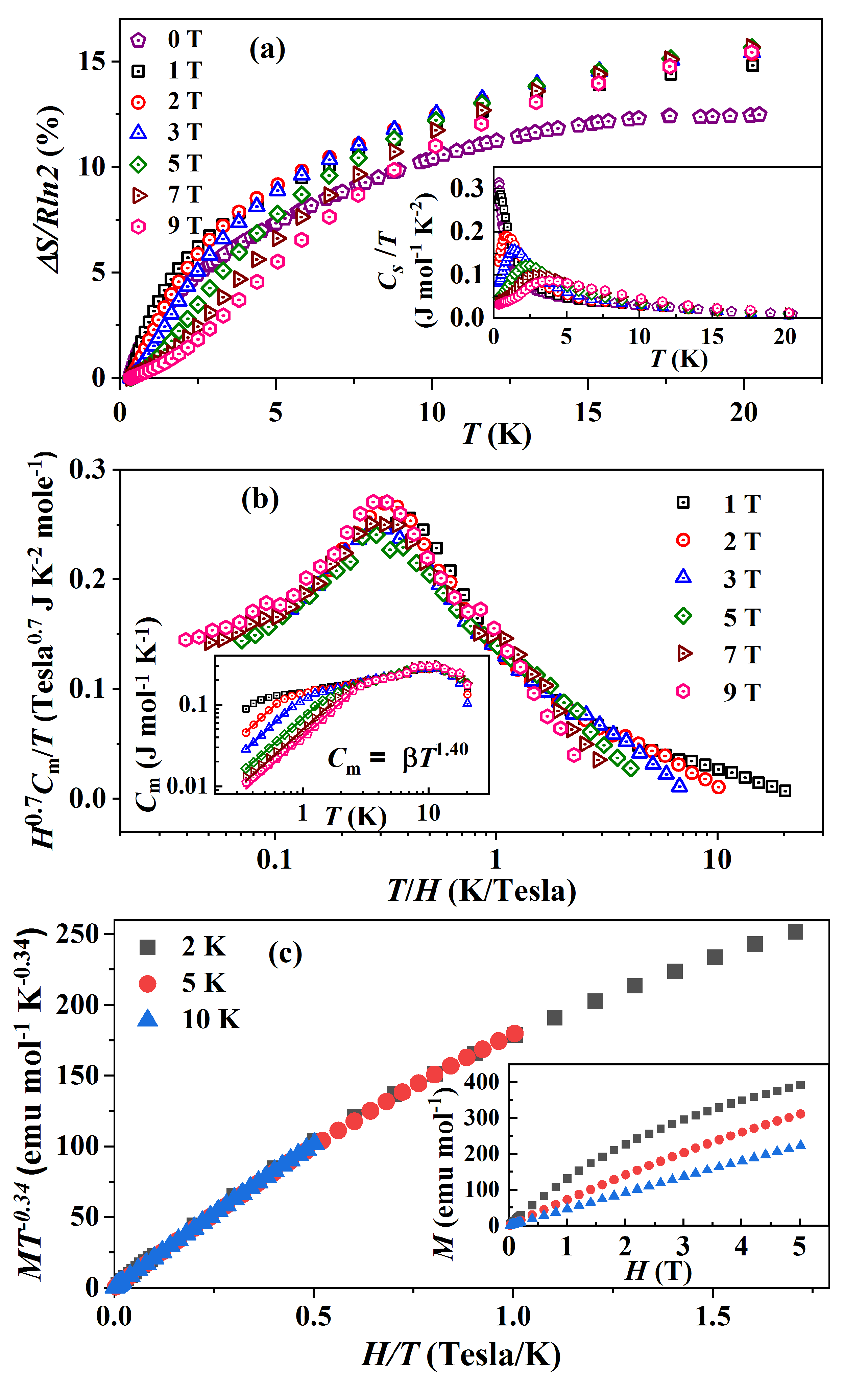}

\centering{}\caption{\label{fig:merged-heat-capacity}(a) The change in the spin entropy, $\Delta S$  as a function of temperature
between 350 mK and 20 K. In presence of a magnetic field, marginally more entropy is released indicating possible partial lifting of frustration by the Zeeman field - another canonical signature of cooperative magnets such as spin-ice.\citep{inbook} Inset shows the spin-contributions to the specific heat, ($C_s = C_p - C_{lat}$)$/T$ {\it vs}. $T$ over the same range. (b) The scaled magnetic heat capacity, $\mathrm{\mathrm{\mathit{C_{\mathrm{m}}}}}H^{\gamma}$$/T$, of YCTO  is plotted against the scaled temperature $T/H$ for various applied fields $H$. The data collapse in the low-temperature regime is consistent with the q = 0 form of the universal function of Ref. \citep{Kimchi2018} which is expected in absence of spin-orbit coupling; inset shows the $\mathrm{C\mathrm{_{m}}}$ of YCTO after deduction of the lattice and Schottky contributions. (c) The M(H) isotherm has been scaled in the form of $M$$T^{\alpha}$ vs. $H$/$T$.}
\end{figure}

Due to the site-dilution, YCTO is similar to doped semiconductors \cite{bhatt1982scaling,potter2012quantum,voelker2001multiparticle} than the recently discovered YbMgGaO$_4$ \cite{li2015gapless}. However, in contrast to the low density of magnetic moments in doped semiconductors, YCTO has a dense ($50\%$) concentration of spins. In the absence of any magnetic order, as established experimentally,  the  natural option for the system in such circumstances is to locally minimize energy of the  antiferromagnetic exchanges by forming singlets. In the process of formation of these singlets, some spins, fewer in number, are left over due to lack of partners. However these spins, sitting on a random network, are not isolated because of the high density of the magnetic ions with which they interact. Indeed, if the background network of the dimers is dynamic, the positions of such unpaired spins are not even static \cite{singh2010valence}. These unpaired dynamic {\it quasispins} \cite{PhysRevLett.111.157201} then interact with each other with effective exchange interactions of the form $\mathcal{H}_{eff}=\sum_{ij} \mathcal{J}_{ij} {\bf S}_i\cdot {\bf S}_j$ where the effective couplings are expected to be $|\mathcal{J}_{ij}|\sim e^{-|{\bf r}_i-{\bf r}_j|/\xi}$ where $\xi$ is the underlying spin correlation length \cite{Kimchi2018a}. Thus, $\mathcal{J}_{ij}$ are weak and random and the fate of the system crucially depends on their distribution as well as the sign structure. Owing to the lack of bipartite structure of the underlying triangular motif, these exchanges are expected to be a mixture of ferromagnetic and antiferromagnetic interactions  eventually leading to a spin-glass state at a much lower temperature \cite{Kimchi2018a,PhysRevLett.75.4302,bhatt1982scaling} depending on the magnitude and distribution of $\mathcal{J}_{ij}$. A power law distribution $\sim \mathcal{J}^{-\gamma}$ leads to a magnetic specific heat scaling, at finite magnetic field of $C_m\sim TH^{-\gamma}$ for $T/H<1$.

The heat capacity measurements reveal a power-law behavior with the universal scaling of the $H^{\gamma}C_{\mathrm{m}}/T$ with $T/H$ and $\gamma\simeq0.7$ (Fig. \ref{fig:merged-heat-capacity}b).  Indeed, recently such a scaling has been argued to result from an intermediate temperature random singlet phase  in a bond disordered system \citep{Kimchi2018}\footnote{In Ref. \citep{Kimchi2018}, data collapse has been shown for the
as-measured specific heat data (without any subtraction of lattice or nuclear/electron Schottky contribution) of Herbertsmithite with $\gamma$ = 0.5. Plotting our data (see Fig. 4 in SM \citep{SM-YCTO}) in this manner is consistent with $\gamma$ = 0.5}. The above conclusions are consistent with the finite dynamic rate observed in muon spin-relaxation experiments where the rate is fairly insensitive to the applied small external magnetic fields.

Further evidence of the correlated nature of the low temperature paramagnet comes from the dependence of the magnetization with magnetic field at low temperatures (inset of Fig. \ref{fig:merged-heat-capacity}c) which cannot be described by a free spin Brillouin function. Indeed, the magnetization after the removal of the contribution of
the free spin magnetization, as estimated in the analysis of the specific heat, shows a scaling collapse of the form $MT^{-\alpha}$ (with $\alpha=0.34\pm 0.02$) as a function of $H/T$ (see Fig. \ref{fig:merged-heat-capacity}c) as expected for the above-mentioned power-law distribution of the effective exchanges with $\alpha=1-\gamma$ \cite{singh2010valence, Kimchi2018}.

In summary, through a complementary set of experiments on the randomly 50\% depleted
triangular lattice $S=1/2$ magnet-- Y$_2$CuTiO$_6$, we establish absence of magnetic order and/or   spin freezing down to
50 mK. This is $0.037\%$ of the Curie-Weiss scale of about $-134$ K, the latter implying strong antiferromagnetic couplings between the magnetic Cu$^{2+}$ spins. While, at even lower temperatures spin-freezing may occur, such drastic suppression of freezing compared to the Curie-Weiss scale opens up a cooperative paramagnetice regime at least between $50$ mK and $2$ K. In the cooperative paramagnet, a scaling collapse, consistent with the random singlet phenomenology \cite{Kimchi2018}, is observed in the magnetic field dependent specific heat and magnetisation.
An exciting question, fuelled by our experimental observation of dynamical signatures, pertains to the role of quantum coherence in the cooperative paramagnet and in particular whether it can support non-trivial entanglement expected in a QSL. While our existing understanding in $\it clean$ frustrated magnets indicates that cooperative paramagnets provide the right background to look for QSLs, search for such quantum coherence in Y$_2$CuTiO$_6$ is clearly a very interesting future step in exploring $\it disorder$ $\it driven$ QSLs. The issue of doping away from the 50\% dilution, or with carrier doping forms similar sets of interesting open questions.

\paragraph*{Acknowledgement :}

Authors thank Kedar Damle for useful discussion. Author groups from IIT Bombay and IISc acknowledge support of measurement facilities in their respective institutions.  Experiments at the ISIS Neutron and Muon Source
were supported by a beamtime allocation RB1900000 from the Science
and Technology Facilities Council. AH thanks Department of Science and Technology, Government of India for support.
AH also acknowledges Newton fund for supporting the visit to ISIS, UK. This work is partly based on experiments
performed at the Swiss Muon Source, Paul Scherrer Institute,
Villigen, Switzerland. SP acknowledges the support of IRCC, IIT Bombay (17IRCCSG011) and SERB, DST, India (SRG/2019/001419). SB acknowledges funding supports through the Department of Atomic Energy, Government of India, under project no. 12-R\&D-TFR-5.10-1100, the Max Planck
Partner group on strongly correlated systems at ICTS and SERB-DST
(India) through project grant No. ECR/2017/000504. DCJ and RM thank the Stiftelsen Olle
Engkvist Byggmaestare and the Swedish Research Council (VR).
AVM, DDS, SP and SB acknowledge hospitality and support
of ICTS and APCTP during the 2nd Asia Pacific Workshop on Quantum
Magnetism (Code: ICTS/apfm2018/11). DDS thanks Science and Engineering Research Board, Department of Science and Technology, Government of India and Jamsetji Tata Trust for support of this research.


\bibliography{Manuscript}

\end{document}


\title{Supplemental Material\\{\it Signatures of a spin-$\frac{1}{2}$ cooperative paramagnet in the diluted triangular lattice of
Y$_{2}$CuTiO$_{6}$}}
\author{S. Kundu}
\thanks{Equal contribution authors}
\affiliation{Department of Physics, Indian Institute of Technology Bombay, Powai,
Mumbai 400076, India}
\author{Akmal Hossain}
\thanks{Equal contribution authors}
\affiliation{Solid State and Structural Chemistry Unit, Indian Institute of Science,
Bengaluru 560012, India}
\author{Pranava Keerthi S}
\affiliation{Solid State and Structural Chemistry Unit, Indian Institute of Science,
Bengaluru 560012, India}
\author{Ranjan Das}
\affiliation{Solid State and Structural Chemistry Unit, Indian Institute of Science,
Bengaluru 560012, India}
\author{M. Baenitz}
\affiliation{Max Planck Institute for Chemical Physics of Solids, 01187 Dresden,
Germany}
\author{Peter J. Baker}
\affiliation{ISIS Pulsed Neutron and Muon Source, STFC Rutherford Appleton Laboratory,
Harwell Campus, Didcot, Oxfordshire OX110QX, UK}
\author{Jean-Christophe Orain}
\affiliation{Paul Scherrer Institute, Bulk MUSR group, LMU 5232 Villigen PSI, Switzerland}
\author{D. C. Joshi}
\affiliation{Department of Engineering Sciences, Uppsala University, Box 534, SE-751 21 Uppsala, Sweden}
\author{Roland Mathieu}
\affiliation{Department of Engineering Sciences, Uppsala University, Box 534, SE-751 21 Uppsala, Sweden}
\author{Priya Mahadevan}
\affiliation{S. N. Bose National Center for Basic Sciences, Block-JD, Salt lake, Kolkata-700106, India}
\author{Sumiran Pujari}
\affiliation{Department of Physics, Indian Institute of Technology Bombay, Powai,
Mumbai 400076, India}
\author{Subhro Bhattacharjee}
\affiliation{International Centre for Theoretical Sciences, Tata Institute of Fundamental
Research, Bengaluru 560089, India}
\author{A. V. Mahajan}
\email{mahajan@phy.iitb.ac.in}
\affiliation{Department of Physics, Indian Institute of Technology Bombay, Powai,
Mumbai 400076, India}
\author{D. D. Sarma}
\email{sarma@iisc.ac.in}
\affiliation{Solid State and Structural Chemistry Unit, Indian Institute of Science,
Bengaluru 560012, India}
\date{\today}

\maketitle

\pagebreak

\subsection{Details of sample preparation and crystal structure of YCTO}
\begin{figure}
\begin{centering}
\includegraphics[scale=0.4]{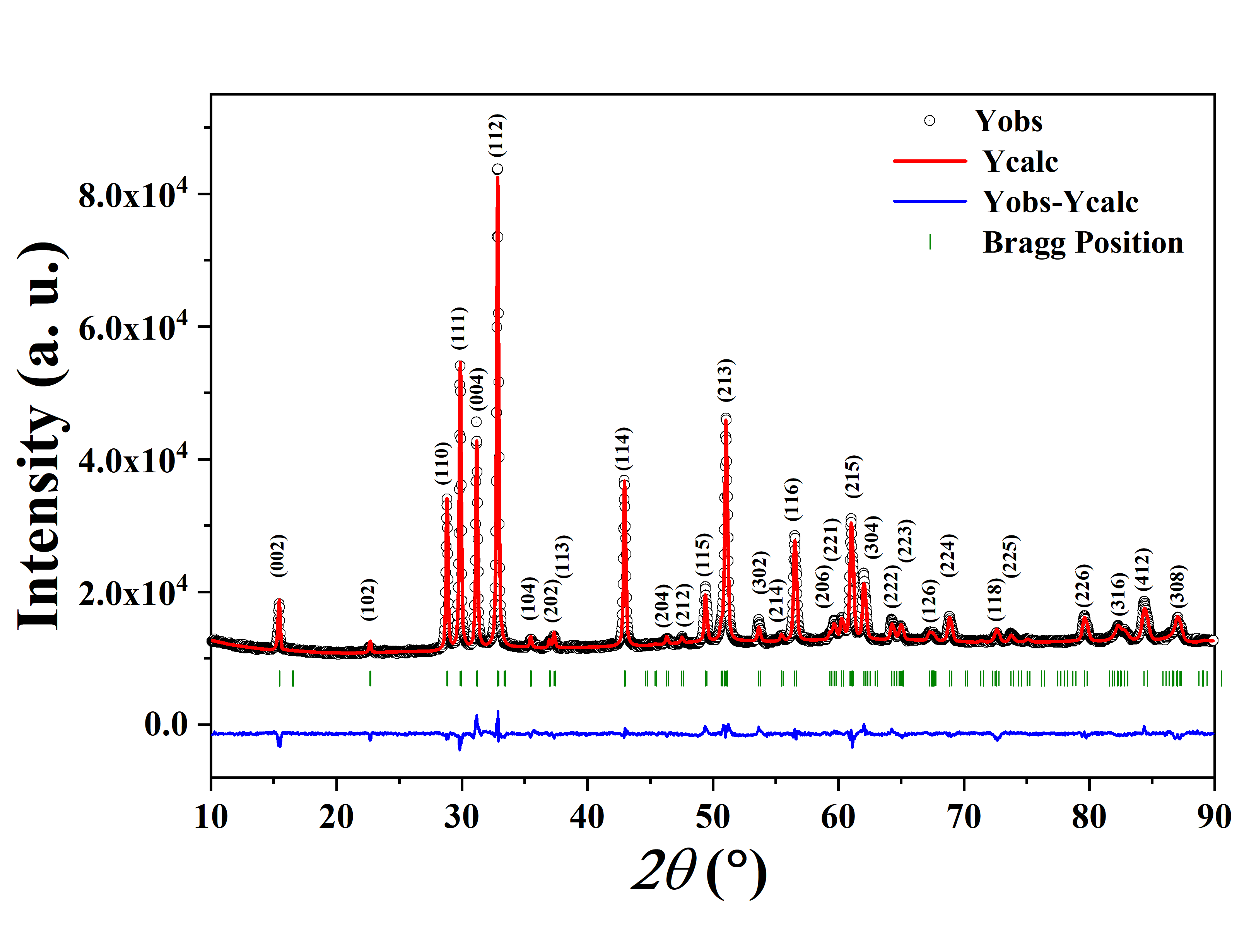}
\par\end{centering}
\caption{\label{fig:refinement of YCTO}XRD refinement of $\mathrm{YCTO}$
(color online). Black circle is observed data, red line is theoretically
calculated, blue line is the difference between the observed and calculated
data and the vertical marks are the Bragg's positions. The plane of
reflection (hkl) is marked for each corresponding Bragg peak. }
\end{figure}

\begin{widetext}
At first, the intimate stoichiometric mixture of $\mathrm{Y_{2}O_{3}}$
(which was preheated at $\mathrm{1150 ^{o}C}$ for 6 hours to remove
carbonate contamination and moisture), $\mathrm{CuO}$ (99.9\%) and
$\mathrm{TiO_{2}}$ (99.9\%) was prepared. Then this mixture was pressed
into a pellet and placed in an alumina crucible and heated in air at $\mathrm{900 ^{o}C}$
for 24 hours in a tubular furnace. Further, the sample was again ground, pelletized
and reheated in air at $\mathrm{1050 ^{o}C}$ for 150 hours with three intermediate
stages of regrinding for better homogeneity.
In the next step, the sample was fired again in air at $\mathrm{1300 ^{o}C}$ for 30 hours to form the final product.

The powder x-ray diffraction data (PXRD) of polycrystalline YCTO sample was collected
at room temperature with Cu-$\mathrm{K_{\alpha}}$ as the radiation source ($\mathrm{\lambda}$ = 1.54056 {\AA})
using Panalytical X-ray diffractometer (Fig. \ref{fig:refinement of YCTO}). The PXRD pattern confirms the formation of single phase YCTO.
After refinement using FullProf software, we found that the prepared YCTO crystallizes in the hexagonal non-centrosymmetric space group $\mathrm{\mathit{P}6_{3}\mathit{cm}}$, as reported earlier \citep{Floros2002}, with lattice parameters
$\mathit{a}$ = $\mathit{b}$ = 6.1951(1) {\AA}, $\mathit{c}$ = 11.4713(2) {\AA}, $\mathrm{\alpha=\beta=90^{o}}$,
$\mathrm{\gamma=120^{o}}$ and is isostructural with $\mathrm{\mathrm{LuMn}O_{3}}$ \citep{Floros2002,Choudhury2010}.
The atomic coordinates of different elements after refinement are
shown in Table \ref{tab:Atomic-coordinates-of YCTO}. The goodness
of the Rietveld refinement as defined by the following parameters
are $\mathrm{\mathit{R}_{p}}$: 12.7\%; $\mathrm{\mathit{R}}_{\mathrm{wp}}$:
8.79\%; $\mathrm{\mathit{R}}_{\mathrm{exp}}$: 4.19\%; $\mathrm{\mathit{\chi}^{2}}$: 4.40.

\begin{table}[h]
\caption{\label{tab:Atomic-coordinates-of YCTO}Atomic coordinates of different
elements of YCTO after refinement.}

\bigskip{}

\centering{}%
\begin{tabular}{cccccc}
\hline
Atom  & Wyckoff position  & x  & y  & z  & Occupancy\tabularnewline
\hline
\hline
Cu  & 6c  & 0.3270  & 0.0000  & -0.0196  & 0.50\tabularnewline
Ti  & 6c  & 0.3270  & 0.0000  & -0.0196  & 0.50\tabularnewline
Y1  & 4b  & 0.3333  & 0.6667  & 0.2196  & 1.00\tabularnewline
Y2  & 2a  & 0.0000  & 0.0000  & 0.2455  & 1.00\tabularnewline
O1  & 6c  & 0.3156  & 0.0000  & 0.1489  & 1.00\tabularnewline
O2  & 6c  & 0.6541  & 0.0000  & 0.3203  & 1.00\tabularnewline
O3  & 2a  & 0.0000  & 0.0000  & -0.0229  & 1.00\tabularnewline
O4  & 4b  & 0.3333  & 0.6667  & 0.4922  & 1.00\tabularnewline
\hline
\end{tabular}
\end{table}
\end{widetext}

\subsection{Experimental details}
\begin{widetext}
Magnetization, $M$, measurements as a function of applied field $H$
(0 to 5 T) and temperature $T$ (in the range 1.8 K to 400 K))
were performed using a Quantum Design SQUID VSM. Zero-field cooled
(ZFC) and field cooled (FC) magnetization measurements in a low field
of 50 Oe were performed down to 500 mK. The heat capacity $C_{\mathrm{p}}(T)$
was measured with a Quantum Design PPMS in various applied fields
down to 0.35\,K. Local probe NMR measurements
were performed on the $^{89}$Y nucleus in a fixed field of 93.95
kOe.  Muon spin relaxation ($\mu$SR) measurements down to 50 mK were performed on the MuSR instrument at ISIS, RAL in UK (for data, see
Ref. \citep{MrRANJANDAS2018}) and PSI in Switzerland.
\end{widetext}

{\subsection{Calculation of the magnetic exchange interaction strengths}

Ab-initio electronic structure calculations were carried out within a plane-wave projected augmented wave \cite{blochl1994projector} implementation of density functional theory in the
Vienna Ab-initio Simulation package (VASP) \cite{kresse1999ultrasoft,kresse1996efficiency}.
The GGA approximation \citep{perdew1996generalized} to the exchange correlation functional was used, in addition
to treating electron-electron interactions at the Cu site with a $U$ of 8 eV \cite{imada1998metal,macridin2005physics,mizuno1998electronic}
within the Dudarev \citep{dudarev1998electron} implementation. A cutoff energy of
600 eV was used to determine the number of plane waves included in the basis. All $k$-space integrations were carried out using
a $\Gamma$ centered 5$\times$5$\times$3 Monkhorst-Pack $k$-point mesh.
A unit cell of YCTO consisting of 3 formula units as shown in Fig. 1 (main manuscript)  with a total of 30 atoms was used for the calculations.
The calculations were carried out for the ferromagnetic (FM) configuration,  an
inter-planar antiferromagnetic configuration (AAFM) where the spins in the plane are ferromagnetic while the coupling between spins in different planes is
antiferromagnetic, as well as AFM where the spins are antiferromagnetically coupled in a plane.
Changing the cutoff energy to 750 meV, changed the energy difference between $E_{FM}$ and E$_{AFM}$ by 0.11 meV ($\sim$0.6\% change), while
increasing the k-points to 9$\times$9$\times$5 changed it by 0.9 meV ($\sim$5\% change), indicating that the results were well converged.
The magnetic part of the total energy, determined for each configuration, was mapped onto a Heisenberg model \citep{zhong2009first} given by
\begin{equation}
{{\cal H}_{mag}} = -\sum\limits_{i,j} {{J_{ij}}{{\bf{e}}_i} \cdot {{\bf{e}}_j}}
\label{eq:classical_heisenberg}
\end{equation}

\noindent where $i,j$ are the site indices, ${J}_{ij}$ denotes the exchange coupling strength between the spins
and ${\bf{e}}_{i,j}$ represents the unit vectors along the corresponding spin directions.
The energy difference ($\Delta$E) in table \ref{tab:energies-YCTO} is defined as $E_{AAFM}-E_{FM}$ and $E_{AFM}-E_{FM}$ for AAFM and AFM, respectively.
Using the energy differences as shown in the table \ref{tab:energies-YCTO} we have estimated the nearest neighbour exchange coupling, $J_{nn}$ and inter-planar exchange coupling, $J_{c}$ to be $\sim -2.90$ meV ($\equiv -33.6$ K) and $\sim -0.09$ meV ($\equiv -1.0$ K), respectively.
A comparison with iso-structural YMnO$_3$ reveals that the ratio of $J_{nn}$ to $J_c$  (= 33.0) in the present case represents an even more two-dimensional nature than that (= 14.2) \citep{zhong2009first} for YMnO$_3$.

\begin{table}
\caption{\label{tab:energies-YCTO}Energetics for different spin configurations.}
\bigskip{}

\centering{}%
\begin{tabular}{ccccc}
\hline
State  & $\Delta$E (meV)\tabularnewline
\hline
\hline
AAFM  & -0.70\tabularnewline
AFM  & -17.74\tabularnewline
\hline
\end{tabular}
\end{table}

The exchange interaction strengths were used to determine the Curie-Weiss temperature which was found to be $\approx -104$ K using the method outlined in \citep{solovyev2012magnetic,nath2014magnetic}, in reasonable agreement with the experimentally obtained value of $\approx -134$ K.}

\subsection{Details of analyses}

\subsubsection*{$^{89}$Y NMR spectra:}
\begin{widetext}

A local probe of the nature of the Cu moments down to 4 K, specifically pertaining to their dynamical/static nature is obtained from the nuclear magnetic resonance of the nearby Y sites, making use of the natural abundance of
the $\mathrm{^{89}Y}$ nuclei with a spin $I=1/2$ and gyromagnetic ratio
$\frac{\gamma}{2\pi}$ = 2.08583 MHz/Tesla.
The variation of the lineshape with $T$ is shown
in Fig. \ref{fig:merged-nmr}a.
While the spectra are asymmetric due to the presence of  different local magnetic environments in this disordered system, the line-width
of the spectrum increases monotonically with decreasing temperature, essentially following the Curie-Weiss susceptibility of the Cu moments and showing no sign of
any divergence down to 4 K.
In order to quantify the line-width and the peak position as a function of the temperature,
\begin{figure}
\includegraphics[scale=0.4]{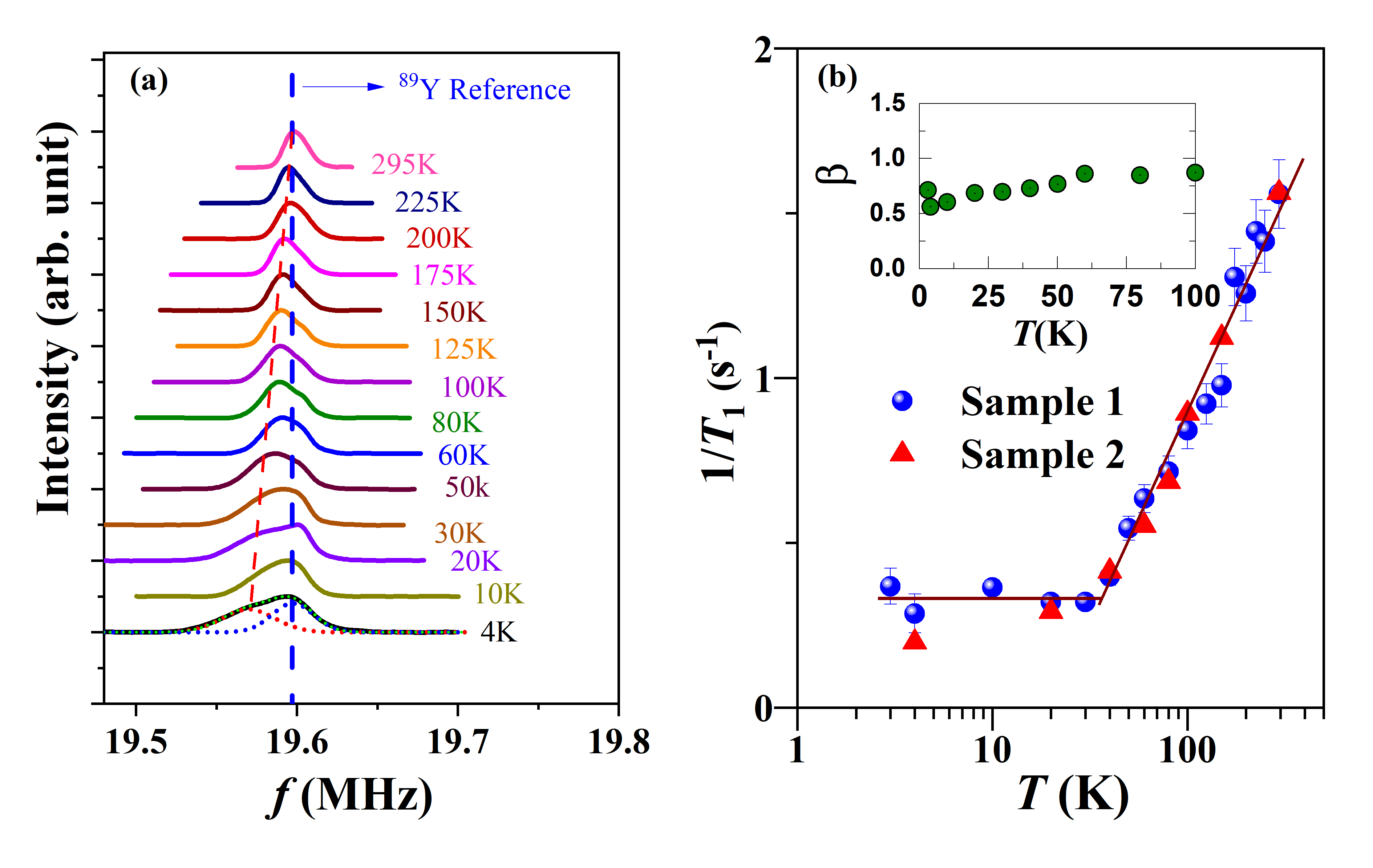}
\caption{\label{fig:merged-nmr}(a) $\mathrm{^{89}Y}$ NMR spectra at different temperatures (the vertical dashed line is the reference frequency for $\mathrm{^{89}Y}$ NMR). A representative two-Gaussian fit is shown for the spectrum at 4 K, (b)
Temperature dependence of the $^{89}$Y spin-lattice relaxation rate $1/T_{1}$ is shown. Inset shows the variation of the stretching exponent with temperature.}
\end{figure}
we minimally fit
the $^{89}$Y spectra at various temperatures to a sum of two Gaussians.
We find that one of the two peaks
remains nearly at the reference
position, while the other shifts gradually  to lower frequencies as temperature
decreases, as also evident in the spectra shown in Fig. \ref{fig:merged-nmr}a.
Specifically, the absence of any divergence/peak in the observed $^{89}$Y NMR shift
and width indicates that the Cu moments remain dynamic down to 4
K.


The $^{89}$Y NMR spin-lattice relaxation rate, 1/$T_{1}$, was obtained by the standard saturation recovery pulse sequence (with a $\frac{\pi}{2}$ pulse of duration 6 $\mu s$) in a field of 93.95 kOe. The recovery of the longitudinal magnetization was fitted to a stretched
exponential ($\mathit{A}\exp(-\frac{\mathit{t_{\mathrm{d}}}}{\mathit{T}_{\mathit{\mathrm{1}}}})^{\beta}$).
The
stretching exponent $\beta$ becomes essentially $1$ above $100$
K, thereby leading to the expected exponential form for $I=1/2$ nuclei.
Significantly,  a stretched exponential recovery
of the nuclear magnetization has been predicted \citep{Kimchi2018}
for the random singlet phase
due to the distribution of magnetic environments that the nuclei sense.
$1/T_{1}$, shown in Fig. \ref{fig:merged-nmr}b, is found to decrease
continuously with the temperature until about 30\,K and
below 30\,K, it is observed to be nearly constant. Once again, the absence
of any peak, indicating a divergence, in the 1/$T_{1}$ vs $T$ data suggests the absence
of a phase transition to an ordered state with the spins remaining
dynamic down to $4$~K \citep{Dally2014}.

\subsubsection*{$\mu$SR:}

The time-evolution of the muon decay asymmetry has been (see Fig. 3 of main manuscript) modeled
by the equation $A(t)=A_{r}e^{-\lambda(T)t}+A_{bg}$,
where $A_{bg}$ takes into account the muons that stop outside the sample and do not get depolarized, and the 1st term, which reflects the contribution from the muons that are stopped inside the sample, contains the relaxation rate (${\lambda}$)  and the relaxing asymmetry ($A_{r}$) . We observed difference in background contributions between helium cryostat ($A_{bg}$ $\mathit{\sim}$$0.085$) and dilution refrigerator ($A_{bg}$ $\mathit{\sim}$$0.050$), which were kept fixed during the data analysis. It is important to mention that the given background constants do not evolve with time. During the analysis of the muon depolarization at applied longitudinal field (see Fig. 3 of main manuscript), we have not put any constraint in the magnitude of relaxing asymmetry and the background contributions, as the given two quantities vary with applied magnetic field presumably because of the deviation of amount of muons that are hitting the sample as well as the change in the direction of the positrons that are coming out of the sample and going towards the detector with the change in applied magnetic field.

\end{widetext}

\subsubsection*{Heat capacity :}
\begin{figure}
\centering{}\includegraphics[scale=0.15]{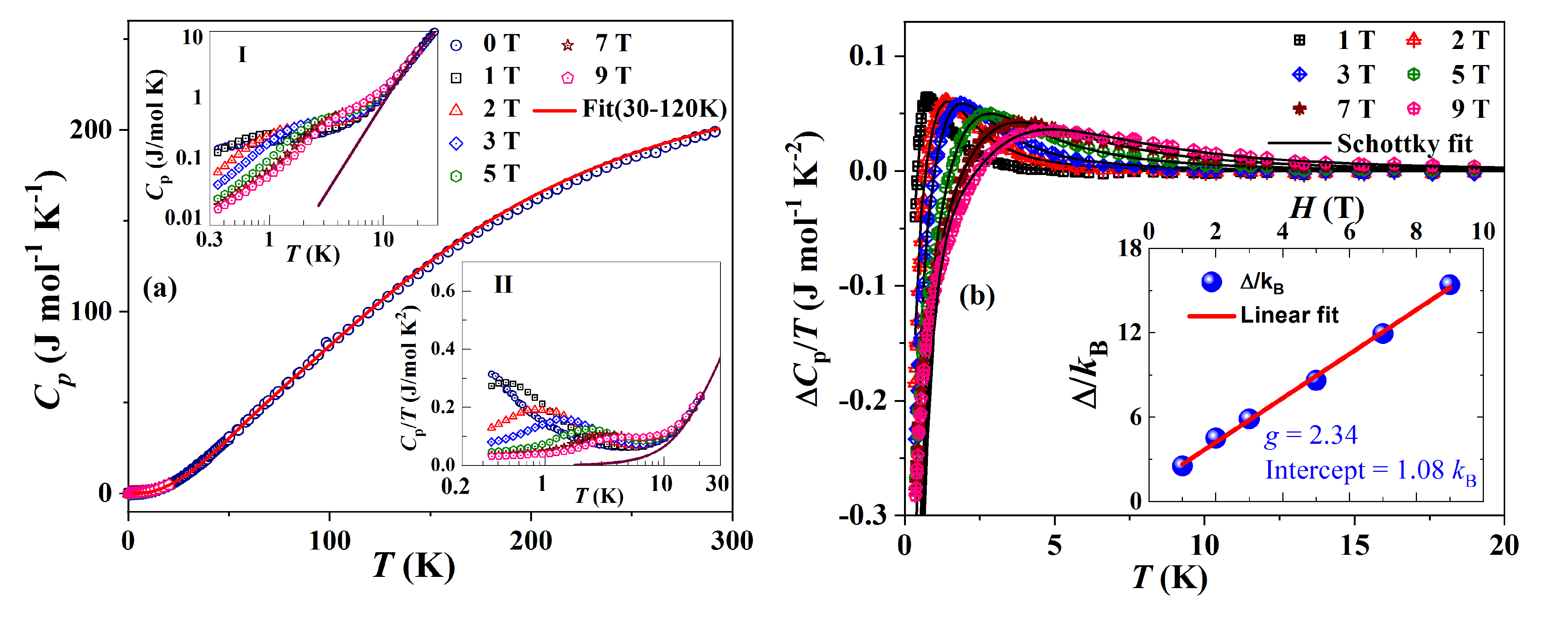}
\caption{\label{fig:Cp vs T YCTO}The main figure in (a) shows the $T$-dependence
of the heat capacity of YCTO in zero field and the
solid line shows the Debye-Einstein fit (see text) of the zero field data. The red solid line is an extrapolation
of the fit to higher and lower temperatures. The inset-I shows an
expanded view of the low-$T$ data in variable magnetic field and
the inset-II shows $C_{\mathrm{p}}/T$ data in the low-$T$ region, (b) Schottky fitting of low temperature anomaly
of YCTO. In the inset, the energy gap ($\Delta/k_{B}$)
as function of applied field is shown and it varies linearly.}
\end{figure}

\begin{widetext}
In zero-field, we measured $\mathit{\mathit{C}_{\mathrm{p}}\mathrm{(}T\mathrm{)}}$
in the $\mathit{T}$-range 0.35 - 290 K (see Fig. \ref{fig:Cp vs T YCTO}a).
It is seen that below about 10 K the $\mathit{\mathit{C}_{\mathrm{p}}\mathrm{(}T\mathrm{)}}$
data is dependent on the applied magnetic field (shown in inset-I
of Fig. \ref{fig:Cp vs T YCTO}a). For clarity we plotted $\mathit{C_{\mathrm{p}}/T}$
vs. $\mathit{T}$ in inset-II of Fig. \ref{fig:Cp vs T YCTO}a. The
low temperature anomaly in $\mathit{\mathit{C}_{\mathrm{p}}/T}$ data
gradually shifts towards the higher temperatures as field increases. This
is nothing but the so called Schottky anomaly. This Schottky contribution ($\mathit{C_{\mathrm{Sch}}}$) to the specific heat most likely arises
from the isolated paramagnetic Cu$^{2+}$ spins or orphan spins within
the system. These spins give rise to a two-level energy system in
the presence of an applied magnetic field. As our specific heat $\mathit{\mathit{C}_{\mathrm{p}}(T)}$
is weakly field dependent at low temperature, we can write the total
specific heat of the system as:
\begin{equation}
C_{\mathrm{p}}(T,H)=C_{\mathrm{lat}}(T)+C_{\mathrm{Sch}}(T,H)+C_{\mathrm{m}}(T,H)\label{eq:Total HC}
\end{equation}
Here, $\mathit{C_{\mathrm{lat}}}$ is the lattice specific heat, $\mathit{C_{\mathrm{Sch}}}$
is the Schottky contribution due to isolated paramagnetic spins within
the system and $\mathit{C_{\mathrm{m}}}$ is the intrinsic magnetic
contribution to the specific heat. While the total spin contribution, $C_s$, from all spins to the total specific heat, $C_p$, can be easily estimated from $C_p - C_{lat}$,  our prime intention is to find
the intrinsic magnetic contribution $\mathit{C_{\mathrm{m}}}$. To
obtain $\mathit{C_{\mathrm{m}}}$, we first have to evaluate $\mathit{C_{\mathrm{lat}}}$
and $\mathit{C_{\mathrm{Sch}}}$ and then subtract these from the
total specific heat $\mathit{C_{\mathrm{p}}}$.
\end{widetext}

\begin{widetext}
In the absence of a suitable non-magnetic analog for YCTO, we used
a combination of Debye and Einstein terms as expressed by Eq.
\ref{eq:Debye} to estimate the lattice contribution $\mathit{C_{\mathrm{lat}}}$.
Here in Eq. \ref{eq:Debye}, $\mathit{C_{\mathrm{D}}}$ and $\mathit{C_{\mathrm{E_{i}}}}$ indicate
the weightage factors corresponding to acoustic and optical modes
of atomic vibrations and $\theta_{\mathrm{D}}$, $\theta_{\mathrm{E}_{\mathrm{i}}}$
are the corresponding Debye and Einstein temperatures, respectively.

\begin{equation}
C_{\mathrm{Debye}}(T)=C_{\mathrm{D}}\left[9R(\frac{T}{\theta_{\mathrm{D}}})^{3}\intop_{0}^{x_{\mathrm{D}}}\frac{x^{4}e^{x}}{(e^{x}-1)^{2}}dx\right]\label{eq:Debye}
\end{equation}
\begin{equation}
C_{\mathrm{Einstein}}(T)=\sum C_{\mathrm{E_{i}}}\left[3R(\frac{\theta_{\mathrm{E}_{i}}}{T})^{2}\frac{exp(\frac{\theta_{\mathrm{E}_{i}}}{T})}{(exp(\frac{\mathrm{\theta}_{\mathrm{E}_{i}}}{T})-1)^{2}}\right]\label{eq:Einstein}
\end{equation}
We fitted the $\mathit{C_{\mathrm{p}}\mathrm{(}T\mathrm{)}}$ data
at 0\,T with Eq. \ref{eq:Debye} for a limited range 30 - 120\,K
and then extrapolated the curve to cover the entire temperature range
0.35 - 290 K (as shown in Fig. \ref{fig:Cp vs T YCTO}a). We checked
several combinations of Debye plus Einstein terms and among them the
best fit we obtained is for one Debye plus two Einstein terms with
weightage factors in the ratio: $C_{\mathrm{D}}$:$C_{\mathrm{E_{1}}}$:$C_{\mathrm{E_{2}}}$
= 1:3:6. The sum $C_{\mathrm{D}}$+$\sum C_{\mathrm{E}_{i}}$ is equal
to 10 which is same as the number of atoms per formula unit of YCTO.
The fitting also yields the Debye temperature $\theta_{\mathrm{D}}$ =
(144$\pm1)$\,K and Einstein temperatures to be $\theta_{\mathrm{E}_{1}}$=
(246$\pm1$)\,K, $\theta_{\mathrm{E}_{2}}$= (610$\pm2$)\,K. Thus
we obtained $\mathit{C_{\mathrm{lat}}}$ which will be subtracted
later from the total $\mathit{C_{\mathrm{p}}\mathrm{(}T\mathrm{)}}$.

Now to eliminate the Schottky contribution, we follow a protocol described
here. In the absence of an applied field, $\mathit{C_{\mathrm{p}}\mathrm{(}\mathrm{0},T\mathrm{)}}$,
the $\mathit{C_{\mathrm{p}}}$ data at $\mathit{H}$ = 0 T, contains
lattice ($\mathit{C_{\mathrm{lat}}}(T)$) as well as magnetic ($\mathit{C_{\mathrm{m}}}$$\mathit{\mathrm{(}H,T}$))
part but lacks Schottky contribution. So, we consider $\mathit{\mathrm{C_{\mathrm{p}}(}\mathrm{0},T\mathrm{)}}$
as a reference and subtract it from higher field data $\mathit{C_{\mathrm{p}}\mathrm{(}H,T}$)
to obtain $\Delta C_{\mathrm{p}}$ {[}$\equiv C_{\mathrm{p}}(H,T)-C_{\mathrm{p}}(0,T)${]}.
Now this $\Delta C_{\mathrm{p}}$ represents the Schottky contribution.
In Fig. \ref{fig:Cp vs T YCTO}b, we have plotted $\Delta C_{\mathrm{p}}/T$
vs. $\mathit{T}$ and fitted by Eq. \ref{eq:Schottky eq} which is
the generalized heat capacity expression for a two level system.
\begin{equation}
C_{\mathrm{Sch}}=f\left[R(\frac{\Delta}{k_{\mathrm{B}}T})^{2}(\frac{g_{0}}{g_{1}})\frac{exp(\frac{\Delta}{k_{\mathrm{B}}T})}{[1+(\frac{g_{0}}{g_{1}})exp(\frac{\Delta}{k_{\mathrm{B}}T})]^{2}}\right]\label{eq:Schottky eq}
\end{equation}
where $\mathit{f}$ is fraction of free spins within the system, $\Delta$
is the Schottky gap, R is the universal gas constant, $k_{\mathrm{B}}$
is the Boltzman constant and $g_{0}$ and $g_{1}$ are the degeneracies
of the ground and excited states respectively. Here $\mathit{g_{\mathrm{0}}=g_{\mathrm{1}}=\mathrm{1}}$.
Fig. \ref{fig:Cp vs T YCTO}b shows that the Schottky anomalies are
well fitted for each field. From the fits, we found the percentage
of free $\mathit{S}$ = $\frac{1}{2}$ spins is $\sim$2\%. The
energy gap ($\Delta$) varies linearly with applied fields shown in the
inset of Fig.  \ref{fig:Cp vs T YCTO}b. From the slope of the linear
fit, we obtained the spectroscopic splitting factor $\mathit{g}$
= 2.34. The linear fit has a non-zero intercept ($\simeq$ 1.08\,K)
at zero field which might indicate the presence of an intrinsic field
within the system.
The as-measured heat capacity (without any subtraction)
is plotted in a scaled manner where the data collapse is evident for $\gamma$ = 0.5 as shown in
Fig. \ref{fig:scaled-Cp-without-subtraction}.

\begin{figure}
\includegraphics[scale=0.35]{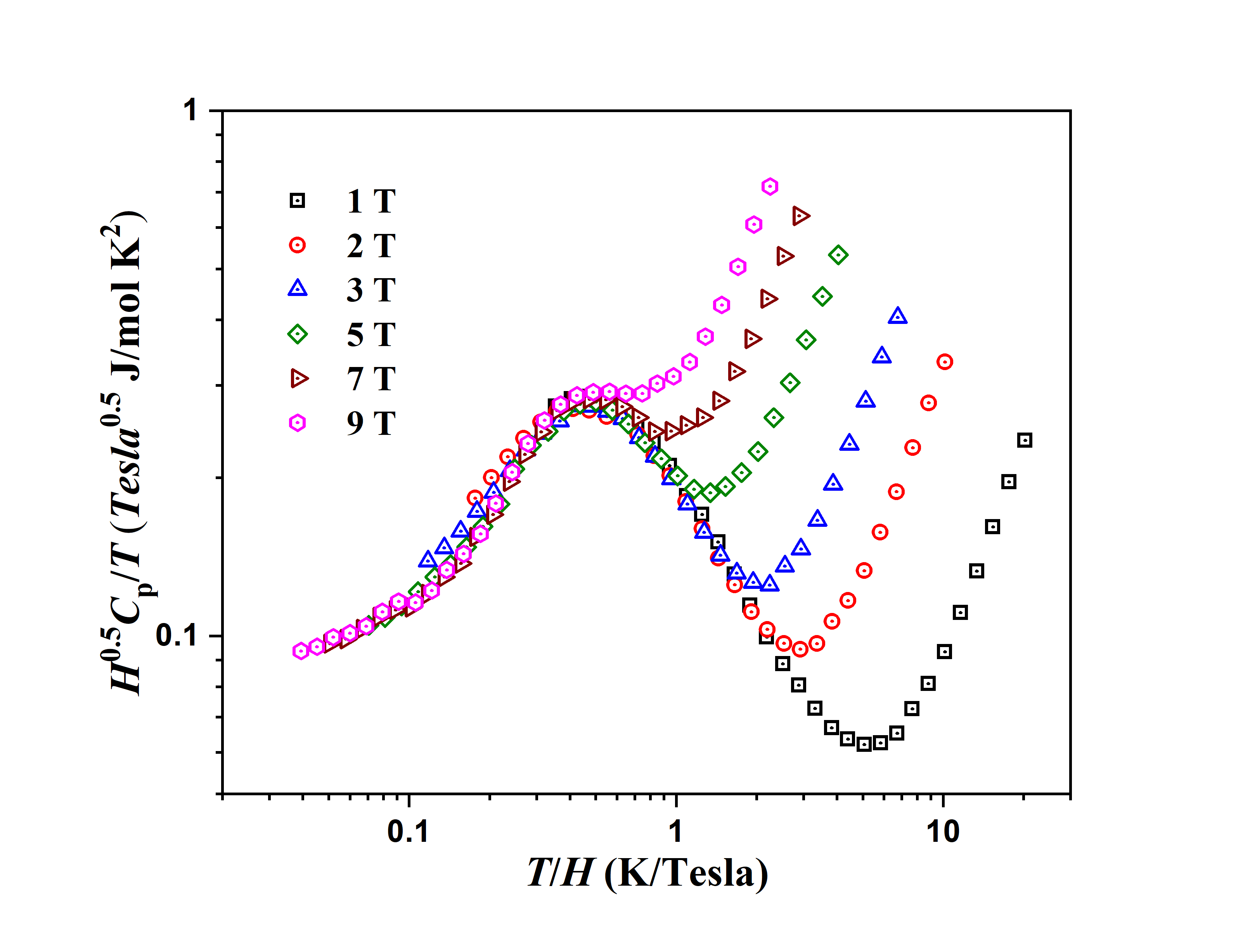}

\caption{\label{fig:scaled-Cp-without-subtraction} The as-measured heat capacity
is plotted as a function of T/H in various fields. The data collapse
is observed with the scaling parameter $\gamma$ = 0.5. }

\end{figure}

\end{widetext}

\subsubsection*{Power law behavior of dc Susceptibility :}
\begin{widetext}

\begin{figure}[h]
\centering{}\includegraphics[scale=0.4]{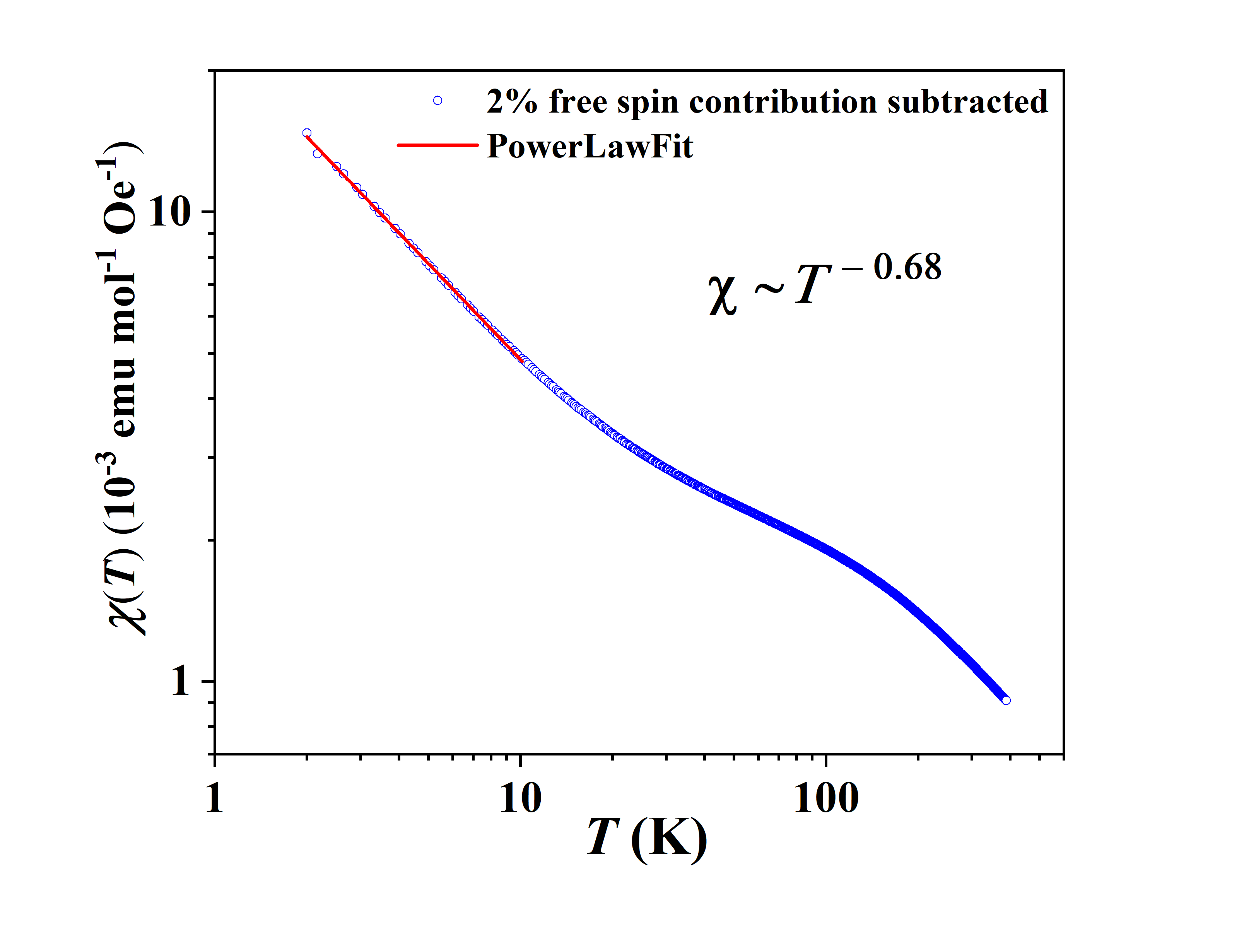}
\caption{\label{fig:Free spin subtracted susceptibility}The susceptibility
as a function of temperature after the subtraction of 2\% free spin
contribution from the measured $\chi(T)$.}
\end{figure}

The fraction of free spins which contribute to the heat capacity through
the Schottky term will also contribute a Curie term to the magnetic
susceptibility, as has been reported in many instances of frustrated magnets. \citep{quilliam2016gapless,nag2016origin,clark2019two,gao2019experimental,choi2019exotic}
If we take the approximate fraction of free spins
to be 2\% and subtract a corresponding Curie term from the measured
dc susceptibility $\chi(T)$ we obtain the data shown in Fig. \ref{fig:Free spin subtracted susceptibility}.
A fit to power law variation at low-$T$ yields $\gamma=0.68$ which
is similar to that obtained from the heat capacity scaling, as expected
for an RSS.
\end{widetext}
\clearpage

\bibliography{Supplemental}